\documentstyle[preprint,aps,version2]{revtex}
\begin{document}

\begin{title}
Probability of Color Rearrangement at Partonic Level \\
in Hadronic $W^+W^-$ Decays 
\end{title}

\author{Zong-Guo Si$^a$, Qun Wang$^a$, Qu-Bing Xie$^{b,a}$}

\begin{instit}
$a$ Department of Physics, Shandong University 

Jinan Shandong 250100, P.R. China

$b$ Center of Theoretical Physics, CCAST(World Lab),

Beijing 100080, P. R. China

\end{instit}

\begin{abstract}
A strict method to calculate the color rearrangement probability at
partonic level in hadronic $W^+W^-$ decays is proposed. The color
effective Hamiltonian $H_c$ is constructed from invariant amplitude for
the process $e^+e^-\rightarrow W^+W^-\rightarrow 
q_1\overline{q}_2q_3\overline{q}_4+ng$~($n=0,~1,~2,~\cdots$) 
and is used to calculate the cross sections of various color singlets and
the color rearrangement probability. The true meaning of the color
rearrangement is clarified and its difference from color interference
is discussed. 

\end{abstract}

\pagestyle{plain}

\section{Introduction}
At LEP2 energy, the real $W^+W^-$ pair production through $e^+e^-$
annihilation becomes possible. More accurate measurements on $W$ mass~($M_W$)
and other properties can be made at this stage of LEP project. 
The success of the precision measurements of $M_W$ relies on 
accurate theoretical knowledge of the dynamics of the production and decay
stages in $e^+e^-\rightarrow W^+W^-\rightarrow
q_1\overline{q}_2q_3\overline{q}_4$. However, the
possible Color Rearrangement~(CR) may obscure the separate identities of
two $W$ bosons so that the final hadronic state may no longer be
considered as a superposition of two separate $W$ decays.
Thus the $W$ mass determination may be distorted by the color
reconnection effect in hadronic $W$-pair decays. This effect was 
first studied by Gustafson, Pettersson and Zerwas~(GPZ)\cite{gus1}. It
attracts a lot of studies in recent years\cite{ts1,ts2,gus,lonn}. 
Considering that the two $W$s decay into two quark pairs 
$q_1\overline{q}_2$ and $q_3\overline{q}_4$, GPZ
assume that these two original color 
singlets can be rearranged into two new ones
$q_1\overline{q}_4$ and $q_3\overline{q}_2$ 
with a probability~($\frac{1}{9}$), and then energetic 
gluons are emitted independently within each new
singlet, which implies that the CR occurs before the parton
shower process begins. But Sj\"ostrand and Khoze do not regard this
instantaneous scenario as a very likely one\cite{ts1,ts2}. The reason
is that the decay vertices of two $W$ bosons are in general separated
in space-time, and therefore the hard gluons~(with $E_g\geq\Gamma_W$)
are produced incoherently by the two pairs $q_1\overline{q}_2$ and
$q_3\overline{q}_4$\cite{ts1,yu}. So there are two color singlets $C_1$
and $C_2$, each containing a $q\overline{q}$ pair and a set of gluons.
Furthermore they argue that the CR in 
Perturbative QCD~(PQCD) phase only comes from the
color interference which should be very small. Hence they conclude that
the non-perturbative contribution dominates the CR effect because the
two color singlets $C_1$ and $C_2$ coexist later 
during the relatively larger space-time scale of hadronization compared with
that of $W^{\pm}$'s life. The non-perturbative CR 
probability is controlled by the space-time overlaps of the color field
induced by two groups of partons in $C_1$ and $C_2$. 
Later on, Gustafson and H\"akkinen stress that CR can only originate
from the partonic level and argue that the hard
gluon emission will enlarge the CR effect.
Because there are increasing ways of color recombination between the
partons of $C_1$ and those of $C_2$ with the growing number of emitted
gluons, even though the rearrangement probability is only
$\frac{1}{N_c^2}$ for each way, the total probability at the partonic
level may be greatly enhanced to the order $\sim (l+1)(m+1)/N_c^2$
where $l$ and $m$ are the numbers of gluons in $C_1$ and $C_2$ respectively.
The final probability can in principle be modified by multiplying
the factor of order $\sim (l+1)(m+1)/N_c^2$ by unknown functions
of variables which characterize the space-time overlaps of the color
fields induced by partons of $C_1$ and $C_2$.
But Gustafson and H\"akkinen's
analysis on the total probability at the partonic level is only a
qualitative one. It does not include many other ways of forming
singlets and the calculation is not based on a strict formulation.

We should keep in mind
that the color fields stretched between partons must be restricted in a
preconfined state, i.e. a singlet. Thus in PQCD phase, CR means the 
transformation from a set of original singlets to that of new ones.
Hence the CR probability from $C_1$~(containing
$q_1\overline{q}_2$) and $C_2$(containing $q_3\overline{q}_4$)
to new recombined ones where $q_1\overline{q}_4$ and
$q_3\overline{q}_2$ belong to different singlets should and can be
estimated by PQCD with more reason and accuracy.
In this paper, we try to calculate the rearrangement probability
from a strict systematic approach of PQCD. This approach is based on
the color effective Hamiltonian\cite{wang} $H_c$ which is built from
the recursive form\cite{bw1,bw2,bw3} of the invariant amplitude $M$ for
the process:
\begin{equation} \label{X}
e^{+}e^{-}\rightarrow W^{+}W^{-}\rightarrow 
q_1\overline{q}_2q_3\overline{q}_4
+ng,~~~~n=0,~1,~2,~\cdots .
\end{equation}
The color effective Hamiltonian $H_c$ is used to calculate 
the cross sections and the probabilities of various rearranged color 
singlets formed by final partons. 
The physical significance of our approach lies in that it includes all
of effects caused by the different space-time intervals between the
decay vertices of two $W$-bosons. The CR probability we obtain shows
that the CR of PQCD stage is not negligible. This seems different from
what Sj\"{o}strand and Khoze conclude in ref.\cite{ts1,ts2}. In our
approach, the meaning of CR can be clearly defined and the difference
from color interference can be easily elucidated. 

The outline of this paper is as follows: in section \ref{ii}, 
we give the invariant amplitude $M_n$ for the 
process~(\ref{X}) in the recursive form; then the color 
effective Hamiltonian $H_c$ is abstracted from $M_n$ in section \ref{iii}; 
thirdly, we use $H_c$ to analyze the color singlet structure of the 
parton states $q_1\overline{q}_2q_3\overline{q}_4+ng$ with $n=0,~1,~2$
and give the rearrangement probabilities in section \ref{iv};
finally, a summary is given in the last section.

\section{Cross Section}
\label{ii}
The differential cross section $d \sigma_n$ for the process~(\ref{X}) is
\begin{equation}
d \sigma_n =\Phi |M_n|^2 d \wp_{n+4}(Q_1, Q_2, Q_3, Q_4, K_1, \cdots, K_n),
\end{equation}
where  $\Phi = \frac{1}{8s}$ is the flux factor multiplied by a spin
average factor, and the phase space factor $d\wp_{n+4}$ is defined by
\[ d \wp_{n+4} = (2\pi)^4 \delta^4(P_+ + P_- - \sum\limits_{i=1}^{4} Q_i
-\sum\limits_{j=1}^{n} K_j) \prod\limits_{i=1}^{4} \frac{d^3
\stackrel{\rightarrow}{Q_i}}{(2\pi)^3 2 Q_i^0} \prod\limits_{j=1}^{n} 
\frac{d^3 \stackrel{\rightarrow}{K_j}}{(2\pi)^3 2 K_j^0}, \]
here $P_+$, $P_-$, $Q_i$~($i=1,~\cdots,~4$) and $K_j$~($j=1,~\cdots,~n$) 
denote the 4-momenta of $e^+$, $e^-$, quarks and gluons respectively.

According to Feynman rules, the matrix element $M_n$ of the
process~(\ref{X})~(see fig.1) can be written as
\begin{equation} \label{m} \begin{array}{rl}
M_n = &\sum\limits_{m=0}^{n} \sum\limits_{V=\nu_e, \gamma^*, Z^0} 
\pounds_{\nu\mu}^{V} D^{\nu\nu'}(W^{+2}) D^{\mu\mu'}(W^{-2}) \\ 
 &\times \hat{S}_{\nu'}(Q_1; K_1, \cdots, K_m;  Q_2) 
\hat{S}_{\mu'}(Q_3;  K_{m+1}, \cdots, K_n; Q_4),
\end{array} \end{equation}
where $W^{\pm}$ denotes the 4-momenta of $W^{\pm}$ boson, and 
$D^{\nu\mu}(W^{\pm 2})$ is the propagator of $W^{\pm}$ boson;
$\pounds_{\nu\mu}^{V}$ is the polarization tensor of leptons; 
$\hat{S}_{\nu}$ is the current containing quarks and gluons which
depends on the 4-momenta, helicities and color indices of the outgoing partons.
In the recursive form, the current $\hat{S}_{\nu}$ can be expressed by 
\begin{equation} \label{s}
\hat{S}_{\nu}(Q_1;K_1, K_2, \cdots, K_m; Q_2) =ieg_s^m 
\sum\limits_{P(1,2,\cdots,m)}
(T^{a_1}T^{a_2}\cdots T^{a_m})^i_j S_{\nu}(Q_1;1, 2, \cdots, m; Q_2),
\end{equation}
where $g_s$ is QCD coupling constant;
$T^a=\frac{\lambda^a}{2}$ and $\lambda^a$ is Gell-Mann
matrix for $SU_c(3)$, and $S_{\nu}$ is the spinor current~(for detail,
see refs.\cite{bw1,bw2}).

Substituting eq.~(\ref{s}) into eq.~(\ref{m}), we obtain
\begin{equation}
\label{mn}
M_n = \sum\limits_{m=0}^{n} 
\sum\limits_{P(1, \cdots, m)}\sum\limits_{P(m+1, \cdots, n)}
(T^{a_1}\cdots T^{a_m})^i_j(T^{a_{m+1}}\cdots T^{a_n})^k_l  
X_{(q_1 g_1 \cdots g_m {\overline{q}}_2)(q_3g_{m+1}\cdots g_n 
\overline{q}_4)},
\end{equation}
where the indices $i,~k$~($j,~l$) denote the color~(anticolor) of outgoing
quark~(antiquark), and 
\[ \begin{array}{ll}
X_{(q_1g_1\cdots g_m \overline{q}_2)(q_3g_{m+1}\cdots g_n \overline{q}_4)} = &
-e^2 g_s^n \sum\limits_{V=\nu_e, \gamma^*, Z^0} 
\pounds_{\nu\mu}^{V} D^{\nu\nu'}(W^{+2}) D^{\mu\mu'}(W^{-2}) \\
 & \times S_{\nu'}(Q_1; K_1, \cdots, K_m; Q_2) 
S_{\mu'}(Q_3; K_{m+1}, \cdots, K_n; Q_4).
\end{array} \]
$\pounds_{\nu\mu}^{V}$ can be written as follows
\begin{equation} \left\{ \begin{array}{lll}
\pounds_{\nu\mu}^{V} &= 
\overline{v}(P_+)[ie\Gamma_{\nu}^{W}]
\frac{iQ_{\alpha}\gamma^{\alpha}}{Q^2} [ie\Gamma_{\mu}^{W}] u(P_-),&V=\nu_e,\\
\pounds_{\nu\mu}^{V}& =\overline{v}(P_+)[ie\Gamma_{\alpha}^{V}] u(P_-)
\tilde{D}^{(V)\alpha\beta}(Q'^2)
[ieF^V_{\beta\nu\mu}(Q',W^+,W^-)],&V=\gamma^*, Z^0, 
\end{array} \right. \end{equation}
where the repetition of indices represents summing~(we use this
convention unless explicitly specified); 
$\Gamma_{\alpha}^{B}$ is the fermion-boson vertices for the boson
$B=\gamma^*,~Z^0,~W^{\pm}$; $Q=W^+ - P_+$ and $Q'= P_+ + P_-$;
$\tilde{D}^{V\alpha\beta}(Q'^2)$ is the propagator of the vector boson 
$V$~(=$\gamma^*$ or $Z^0$); 
$F^V_{\beta\nu\mu}(Q',W^+,W^-)$ is the three-boson vertex of
$\gamma^* W^+W^-$~($V=\gamma^*$) or $Z^0 W^+W^-$~($V=Z^0$).

Up to now, we use the recursive formula to write down the
invariant amplitude $M_n$ for the process $e^+e^-\rightarrow W^+W^-
\rightarrow q_1\overline{q}_2q_3\overline{q}_4+ng$.
In this form, we see that $M_n$ can be clearly expressed
as a uniform color part multiplied by a momentum function of partons.
The effective Hamiltonian $H_c$ can be found from this form of the
amplitude. This is what we shall do in the next section.

\section{Color Effective Hamiltonian $H_c$}
\label{iii}
In ref.\cite{wang}, from PQCD, a strict formulation has been
proposed to calculate the cross section of color singlets for the
process $e^+e^- \rightarrow \gamma^*/Z^0 \rightarrow q\overline{q} +ng$.
Now we use the same approach
to abstract the color effective Hamiltonian $H_c$ for the process~(\ref{X})
from the invariant amplitude $M_n$ given in eq.~(\ref{mn}). 
Then $H_c$ is found in the following form:
\begin{equation}
\label{hc}
\begin{array}{lcl}
H_c &=& \sum\limits_{m=0}^{n} \sum\limits_{P(1, \cdots, m)}
\sum\limits_{P(m+1, \cdots, n)} 
(T^{a_1}\cdots T^{a_m})^i_j(T^{a_{m+1}}\cdots T^{a_n})^k_l
X_{(q_1 g_1 \cdots g_m {\overline{q}}_2)(q_3g_{m+1}\cdots g_n \overline{q}_4)} \\
 & & \times \Psi_{1i}^+ \Psi_2^{j+} \Psi_{3k}^+ \Psi_4^{l+} 
A^{a_1 +} \cdots A^{a_n +} \\
 &=&(\frac{1}{\sqrt{2}})^n \sum\limits_{m=0}^{n}
\sum\limits_{P(1, \cdots, m)}\sum\limits_{P(m+1, \cdots, n)}
[\Psi_{1i}^+ \Psi_2^{j+} (G_1 \cdots G_m)^i_j]
[\Psi_{3k}^+ \Psi_4^{l+} (G_{m+1}\cdots G_n)^k_l] \\
 & & \times 
X_{(q_1 g_1 \cdots g_m {\overline{q}}_2)(q_3g_{m+1}\cdots g_n \overline{q}_4)},
\end{array} \end{equation}
where $\Psi_{ui}^+ = (R^+, Y^+, B^+)_u$ and 
$\Psi_u^{j+} = (\overline{R}^+, \overline{Y}^+, \overline{B}^+)_u$ are 
the color creation
operator for quark and antiquark, respectively. The color octet operator
$G_u$ of gluon $u$ is defined by
\begin{equation} 
G_{uj}^i =\frac{1}{\sqrt{2}} (\lambda^{a_u} A_u^{a_u})_j^i 
    = \Psi_u^{i+} \Psi_{uj}^+ 
    - \frac{1}{3}\Psi_u^{x+} \Psi_{ux}^+\delta_j^i,
\end{equation}
where $A_u^{a_u}$~($a_u=1, \cdots, 8$) is  color operator for gluon $u$
and is defined in ref.~\cite{wang}. 
Here $H_c$ is another expression of $S$ matrix, so it is not
necessarily Hermitian. 

For a final color state $|f>$, its cross section can be calculated by
\begin{equation}
\label{eq9}
\sigma_n^f = \int \Phi |<f|H_c|0>|^2 d\wp_{n+4},
\end{equation}
where $|0>$ is the initial state free of color. If
$$|f>=|[\Psi_{1i}^+ \Psi_2^{j+} (G_1 \cdots G_m)^i_j]
[\Psi_{3k}^+ \Psi_4^{l+} (G_{m+1}\cdots G_n)^k_l]>,$$
after summing over all of the color indices, we have 
\begin{equation} \begin{array}{ll}
\sum\limits_f \sigma_n^f &=
\int\Phi \sum\limits_f |<f|H_c|0>|^2 d\wp_{n+4}
 = \int\Phi <0|H_c^+ H_c|0> d\wp_{n+4} \\
 & =\int \Phi |M_n|^2 d\wp_{n+4} =\sigma_n.
\end{array} \end{equation}
It shows that the calculation of the ordinary cross section via $H_c$
returns to the original form. So the validity of $H_c$ for the 
process~(\ref{X}) is verified.

For the parton system $q_1\overline{q}_2q_3\overline{q}_4+ng$, 
the color state is composed of the color charges of $q_1$, 
$\overline{q}_2$, $q_3$, $\overline{q}_4$ and
$n$ gluons. It belongs to the color space
\[3_1\bigotimes 3_2^* \bigotimes 3_3 \bigotimes 3_4^*
\bigotimes 8_1 \bigotimes \cdots \bigotimes 8_n. \]
There are many ways of reducing this color space. Corresponding to 
each reducible way is one set of orthogonal singlet sub-spaces 
whose bases contribute a complete and orthogonal set of color singlets. 
For a color singlet set
\{$|f_k>,~k=1,~\cdots$\}, we have
\begin{equation}\label{cc}
|f_k><f_k|=1,~~~~~<f_k|f_l>=\delta_{kl} 
\end{equation}
and 
\begin{equation} \label{se} \begin{array}{ll}
\sum\limits_k \sigma_n^k &=
\int\Phi \sum\limits_k |<f_k|H_c|0>|^2 d\wp_{n+4} 
 = \int\Phi <0|H_c^+|f_k><f_k|H_c|0> d\wp_{n+4} \\
 &=\int\Phi <0|H_c^+H_c|0>d\wp_{n+4}=\int\Phi |M_n|^2 d\wp_{n+4}=\sigma_n.
\end{array} \end{equation}
This is the result of unitarity.

According to eq.~(\ref{hc}), 
for instance, one can get the concrete expressions of color effective
Hamiltonian $H_{cn}$ for the 
the process~(\ref{X}) with  $n=0,~1,~2$ as follows:\\
a.) for the process $e^+e^-\rightarrow W^+W^-\rightarrow
q_1\overline{q}_2q_3\overline{q}_4$~($n=0$),
\begin{equation}\label{hc0}
H_{c0} = (\Psi_{1i}^+ \Psi_2^{i+})(\Psi_{3j}^+ \Psi_4^{j+})
X_{(q_1\overline{q}_2)(q_3\overline{q}_4)};
\end{equation}
b.) for the process $e^+e^-\rightarrow W^+W^-\rightarrow
q_1\overline{q}_2q_3\overline{q}_4g_1$~($n=1$),
\begin{equation}\label{c2}
H_{c1} = H_{c1}^1 + H_{c1}^2,
\end{equation}
where 
\[ \begin{array}{ll}
H_{c1}^1 &=\frac{1}{\sqrt{2}}
(\Psi_{1i}^+ G_{1j}^{i} \Psi_2^{j+}) (\Psi_{3k}^+ \Psi_4^{k+}) 
X_{(q_1 g\overline{q}_2)(q_3\overline{q}_4)},\\
H_{c1}^2 &=\frac{1}{\sqrt{2}}
(\Psi_{1i}^+ \Psi_2^{i+}) (\Psi_{3j}^+ G_{1k}^{j} \Psi_4^{k+})
X_{(q_1 \overline{q}_2)(q_3g\overline{q}_4)};
\end{array} \]
c.)  for the process $e^+e^-\rightarrow W^+W^-\rightarrow
q_1\overline{q}_2q_3\overline{q}_4g_1g_2$~($n=2$),
\begin{equation}\label{c3} 
H_{c2} = H_{c2}^1 + H_{c2}^2 + H_{c2}^3,
\end{equation}
where
\[ \begin{array}{ll}
H_{c2}^1 =&\frac{1}{2}
\sum\limits_{P(1,2)}
(\Psi_{1i}^+ G_{1j}^{i} G_{2k}^j \Psi_2^{k+}) (\Psi_{3l}^+ \Psi_4^{l+})
X_{(q_1 g_1 g_2 \overline{q}_2)(q_3 \overline{q}_4)}, \\
H_{c2}^2 =&\frac{1}{2} \sum\limits_{P(1,2)}
(\Psi_{1i}^+ \Psi_2^{i+}) (\Psi_{3j}^+ G_{1k}^j G_{2l}^k  \Psi_4^{l+})
X_{(q_1 \overline{q}_2)(q_3 g_1 g_2 \overline{q}_4)}, \\
H_{c2}^3 =&\frac{1}{2} \sum\limits_{P(1,2)}
(\Psi_{1i}^+ G_{1j}^{i} \Psi_2^{j+}) (\Psi_{3k}^+ G_{2l}^k \Psi_4^{l+}) 
X_{(q_1 g_1 \overline{q}_2)(q_3g_2 \overline{q}_4)}. 
\end{array} \]

\section{Color Singlet Structure of the final parton system}
\label{iv}
In this section, we try to use $H_c$ derived in last section 
to study the color singlet structure of the final parton system, give
the CR probability  and compare our results with those of other
authors. As the gluon number grows, it becomes more and more difficult to
calculate $\sigma_n$, and the 
number of different ways of forming color singlets increase drastically.
To illustrate what our approach is and how it works, 
we study only three lowest order cases: 
$q_1\overline{q}_2q_3\overline{q}_4$, $q_1\overline{q}_2q_3\overline{q}_4g_1$
and $q_1\overline{q}_2q_3\overline{q}_4g_1g_2$. 

A.) For the parton system $q_1\overline{q}_2q_3\overline{q}_4$, 
due to the reasons given in ref.\cite{ts2}, we have no need to consider the
color configurations resulting from the reductions of 
$3\bigotimes 3=3^*\bigoplus 6$ and $3^*\bigotimes 3^*=3\bigoplus 6^*$.
Hence the color space  $3_1 \bigotimes 3_2^* \bigotimes 3_3 \bigotimes
3_4^*$ has only two reducible ways which correspond to the color
configurations with and without CR.
In the following, we will discuss these two cases in detail. 

For the reducible way
\[            (3_1 \bigotimes 3_2^*) \bigotimes (3_3 \bigotimes 3_4^*)
\rightarrow (1_{12}\bigoplus 8_{12}) \bigotimes (1_{34}\bigoplus 8_{34})
\rightarrow (1_{12}\bigotimes 1_{34}) \bigoplus (8_{12}\bigotimes 8_{34}),\]
the color singlet set is $\{|\tilde{f}_0^i>,~i=1,~2\}$ where
\begin{equation}\label{cc1}
|\tilde{f}_0^1> =\frac{1}{3}|(\Psi_{1i}^+\Psi_{2}^{i+})(\Psi_{3j}^+\Psi_4^{j+})>,
~~~~~|\tilde{f}_0^2> =\frac{1}{\sqrt{8}}|Tr(G_{12} G_{34})>,
\end{equation}
where $G_{xyi}^k=\Psi_{xi}^+ \Psi_y^{k+} -\frac{1}{3}\Psi_{xl}^+ 
\Psi_y^{l+}\delta_i^k$~($xy=12$, or $34$) denotes the color octet state 
formed by $q_x$ and $\overline{q}_y$. 
In $|\tilde{f}_0^1>$, 
$\Psi_{1i}^+\Psi_{2}^{i+}$ and $\Psi_{3j}^+\Psi_4^{j+}$ represent the two
initial color singlets within $q_1\overline{q}_2$ and $q_3\overline{q}_4$.
Defining the probabilities of $|\tilde{f}_0^i>~(i=1,~2)$ as 
$\tilde{P}_0^i=\frac{\int\Phi|<\tilde{f}_0^i|H_{c0}|0>|^2 d\wp_4}{\sigma_0}$ 
where $\sigma_0$ is the cross section for 
$e^+e^-\rightarrow W^+W^-\rightarrow q_1\overline{q}_2q_3\overline{q}_4$ and
$H_{c0}$ is given in eq.~(\ref{hc0}), we find that $\tilde{P}_0^1 = 100\%$
and $\tilde{P}_0^2=0$. 
This set corresponds to the original color configurations.

For the other reducible way
\[            (3_1 \bigotimes 3_4^*) \bigotimes (3_3 \bigotimes 3_2^*)
\rightarrow (1_{14}\bigoplus 8_{14}) \bigotimes (1_{32}\bigoplus 8_{32})
\rightarrow (1_{14}\bigotimes 1_{32}) \bigoplus (8_{14}\bigotimes 8_{32}),\]
the color singlet set is $\{|f_0^i>,~i=1,~2\}$ where
\begin{equation}
 |f_0^1> =\frac{1}{3}|(\Psi_{1i}^+\Psi_{4}^{i+})(\Psi_{3j}^+\Psi_2^{j+})>,
~~~~~|f_0^2> =\frac{1}{\sqrt{8}}|Tr(G_{14} G_{32})>.
\end{equation}
Their probabilities are given by
\begin{equation}
\label{eq13}
P_{0}^1=\frac{\int\Phi|<f_0^1|H_{c0}|0>|^2 d\wp_4} {\sigma_0}
=\frac{1}{9},~~~~~
P_{0}^2=\frac{\int\Phi|<f_0^2|H_{c0}|0>|^2 d\wp_4} {\sigma_0}
=\frac{8}{9}.
\end{equation}
One sees that the probability of $|f_0^1>$, where 
$q_1\overline{q}_4$ and $q_3\overline{q}_2$ form two singlets
$\Psi_{1i}^+\Psi_{4}^{i+}$ and $\Psi_{3j}^+\Psi_2^{j+}$,
is $\frac{1}{9}$. This state is just
the CR case first discussed by 
GPZ\cite{gus1}. But here 
it clearly shows that the states with and without CR 
belong to two different completeness sets. In the new set, the 
state $f_0^2$ is the 
color singlet made up of two color octets which are formed from color
charges of the pair $q_1\overline{q}_4$ and $q_3\overline{q}_2$ respectively.
So all of the four partons must in principle hadronize as a unity in this
state, though we do not know how to treat this kind of hadronization
rigorously. Note that the state 
$|f_0^2>$ is different from $|\tilde{f}_0^1>$.
But in LUND string fragmentation picture, in $|f_0^2>$ which is built up
by two octets, the neutral color flow connects 
$q_1$ with $\overline{q}_2$ and $q_3$ with $\overline{q}_4$, and the 
substrings stretched within the pair $q_1\overline{q}_2$ and 
$q_3\overline{q}_4$ are treated as subsinglets~(in fact, at the
partonic level, they are not singlets, but only color neutral objects). 
Then the hadronization result of $|f_0^2>$ certainly has
no difference  with those of $|\tilde{f}_0^1>$ in LUND model. 

B.) The color space
$3_1 \bigotimes 3_2^* \bigotimes 3_3 \bigotimes 3_4^* \bigotimes 8_1$
of the parton system $q_1\overline{q}_2q_3\overline{q}_4g_1$ can be reduced
in two ways: one corresponds to the case with
no CR, and the other to that with CR.
Here we discuss the latter case. 
Corresponding to the following reducible way 
\[ \begin{array}{ll}
& (3_1\bigotimes 3_4^*)\bigotimes (3_3\bigotimes 3_2^*)\bigotimes 8_1
\rightarrow   (1_{14}\bigoplus 8_{14})\bigotimes 
(1_{32}\bigoplus 8_{32}) \bigotimes 8_1 \\
\rightarrow & (1_{14}\bigotimes 8_{32}\bigotimes 8_1) \bigoplus 
       (8_{14}\bigotimes 1_{32}\bigotimes 8_1) 
       \bigoplus (8_{14}\bigotimes 8_{32}\bigotimes 8_1),
\end{array} \]
the completeness set of color singlets is $\{|f_1^j>,~j=1,~\cdots,~4\}$ where
\begin{equation} \label{cc2} \left\{ \begin{array}{llll}
|f_1^1>&=\frac{1}{\sqrt{24}}|Tr(G_{14} G_{1})(\Psi_{3x}^+ \Psi_2^{x+})>,&
|f_1^2>&=\frac{1}{\sqrt{24}}|Tr(G_{32} G_{1})(\Psi_{1x}^+ \Psi_4^{x+})>,\\
|f_1^3> &=\sqrt{\frac{3}{80}} |Tr(\{G_{14}, G_1\} G_{32})>, &
|f_1^4> &=\frac{1}{\sqrt{48}} |Tr([G_{14}, G_1] G_{32})>,
\end{array} \right. \end{equation}
with
\[\left\{ \begin{array}{l}
\{G_1, G_2\}_k^i = G_{1l}^iG_{2k}^l +G_{2l}^iG_{1k}^l 
- \frac{2}{3}Tr(G_1 G_2) \delta_k^i, \\
{[G_1, G_2]}_k^i = G_{1l}^iG_{2k}^l - G_{2l}^i G_{1k}^l.
\end{array} \right. \]
Their probabilities $P_1^i$~($i=1,~\cdots,~4$) are given by
\begin{equation} \label{eq16} \begin{array}{ll}
P_{1}^i&=\frac{1}{\sigma_1}{\int\Phi|<f_1^i|H_{c1}|0>|^2d\wp_5}
=\frac{1}{\sigma_1}{\int\Phi|<f_1^i|H_{c1}^1|0>
+<f_1^i|H_{c1}^2|0>|^2d\wp_5} \\
 &=\frac{1}{\sigma_1}{\int\Phi[\sum\limits_{j=1}^2|<f_1^i|H_{c1}^j|0>|^2
+2Re(<0|H_{c1}^{1+}|f_1^i><f_1^i|H_{c1}^2|0>)]d\wp_5},
\end{array} \end{equation}
where $\sigma_1$ is the cross section of the process
$e^+e^-\rightarrow W^+W^-\rightarrow q_1\overline{q}_2q_3\overline{q}_4g_1$,
and $H_{c1}$ is given in eq.~(\ref{c2}). They are related to the
energy $\sqrt{S}$ and PQCD parameters~(e.g. $\alpha_s$ and $Y_{min}$ etc.).
$\int\Phi\sum\limits_{j=1}^2|<f_1^i|H_{c1}^j|0>|^2d\wp_5$ are
proportional to 
$\sigma_1$. Thus in the last line of eq.~(\ref{eq16}), the first term
does not depend on these quantities. The second term is the color
rearrangement caused by the color interference. Even if it vanishes, the color
rearrangement still exists. This shows that the color interference
contributes to only part of CR. Additionally,
it is easy to verify $\sum\limits_{i=1}^4 P_1^i=1$ which is the natural 
result of eq.~(\ref{se}). 

We notice that in $|f_1^1>$, $Tr(G_{14} G_{1})$ is the color singlet
formed by $q_1$, $\overline{q}_4$ and $g_1$, while $\Psi_{3x}^+\Psi_2^{x+}$
the color singlet by $q_3$ and $\overline{q}_2$. The case
of $|f_1^2>$ is similar to that of $f_1^1$. 
These two states are just the CR states 
discussed by Gustafson and H\"{a}kkinen in their naive
model\cite{gus}. But the meaning of the probabilities of these two
states in their work is different from that in ours.
For the process $e^+e^-\rightarrow W^+W^- \rightarrow
q_1\overline{q}_2q_3\overline{q}_4g_1$, they consider two possible cases:
in one, $g_1$ is radiated from $q_1\overline{q}_2$ with a probability
$\tilde{P}$, and in the other, $g_1$ is radiated from $q_3\overline{q}_4$
with a probability $(1-\tilde{P})$. There are no interference
terms. For each case, they give the CR probability
$\frac{2}{9}$. So according to their analysis, one can derive
that the total CR probability is $\frac{2}{9}$ for the
whole process, since
$\frac{2}{9}\tilde{P}+\frac{2}{9}(1-\tilde{P})=\frac{2}{9}$. 
But in our calculation, the probability is expressed in eq.~(\ref{eq16}).
One cannot distinguish  which source the gluon is radiated
from. Since in eq.~(\ref{c2}), the first
term $H_{c1}^1$ describes the case 
that the gluon is emitted from $W^+$ and the second one $H_{c1}^2$
describes that the gluon is from $W^-$. These two terms
combined to give probabilities in eq.~(\ref{eq16}). 
Note that in this paper, as was done in ref.\cite{gus}, we also only
consider the hard gluon emission. In this case, we
find that the interference terms are negligibly small, and we
obtain $P_1^1=P_1^2\simeq \frac{1}{9}$, $P_1^3\simeq \frac{5}{18}$ and
$P_1^4\simeq \frac{1}{2}$ from eq.~(\ref{eq16}).
We notice that $|f_1^3>$ or $|f_1^4>$ represents the color singlet formed by
two color octets: one is the octet formed by two octets from $g_1$ and 
$q_1\overline{q}_4$ which are symmetric in $|f_1^3>$ or antisymmetric 
in $|f_1^4>$, and the other is from $q_3\overline{q}_2$. 
In this two states, the four quarks are included in a whole
singlet so they do not emerge in two different subsinglets. As we recall
that CR means the two original singlets
$C_1$(containing $q_1 \overline{q}_2$) and $C_2$(containing $q_3
\overline{q}_4$) are rearranged to make $q_1 \overline{q}_4$ and $q_3
\overline{q}_2$ enter into two new different 
subsinglets, our result of CR probability is $2/9$. It is
interesting to further look at how LUND model treats the hadronization of the
states $|f_1^3>$ and $|f_1^4>$. According to the neutral color flow picture
of LUND model, such states result in two neutral color flows: one
from $q_1$ to  $\overline{q}_2$ via $g_1$, and the other from $q_3$ to
$\overline{q}_4$, or one from $q_1$ to $\overline{q}_2$, and the other from 
$q_3$ to $\overline{q}_4$ via $g_1$. These two neutral flows were
approximated to singlet string pieces which fragment into hadrons
independently. So in LUND model, no difference exists of the 
hadronization result of $|f_1^3>$ and $|f_1^4>$ from those of
the states with no CR. In this sense, the CR probability 
is also about $\frac{2}{9}$. 

C.) For the parton state $q_1\overline{q}_2q_3\overline{q}_4g_1g_2$,
to obtain the singlet set that results in the CR, we
reduce its color space $3_1 \bigotimes 3_2^* \bigotimes 3_3 \bigotimes
3_4^*\bigotimes 8_1 \bigotimes 8_2$ as follows:
\begin{equation}\label{my} \begin{array}{ll}
   & (3_1 \bigotimes 3_4^*) \bigotimes (3_3 \bigotimes 3_2^*) 
\bigotimes 8_1 \bigotimes 8_2 
\rightarrow  (1_{14} \bigoplus 8_{14}) \bigotimes (1_{32}
\bigoplus 8_{32}) \bigotimes 8_1 \bigotimes 8_2 \\
\rightarrow & [1_{14}\bigotimes 1_{32}\bigotimes 8_1\bigotimes 8_2]
 \bigoplus [1_{14}\bigotimes 8_{32}\bigotimes 8_1\bigotimes 8_2] 
 \bigoplus [8_{14}\bigotimes 1_{32}\bigotimes 8_1\bigotimes 8_2]\\
 &\bigoplus [8_{14}\bigotimes 8_{32} \bigotimes 8_1\bigotimes 8_2].
\end{array} \end{equation}
Obviously the color singlets corresponding to the first three
terms in the above equation lead to CR. The last
term of eq.~(\ref{my}) can be further reduced as, e.g.\\
$\bf{(a).}$ $(8_{14}\bigotimes 8_1) 
\bigotimes (8_{32}\bigotimes 8_2) \rightarrow 
[1_{8_{14}\bigotimes 8_1}\bigotimes 1_{8_{32}\bigotimes 8_2}]
   \bigoplus [O_{8_{14}\bigotimes 8_1}\bigotimes
O'_{8_{32}\bigotimes 8_2}]$, \\
$\bf{(b).}$ $(8_{14}\bigotimes 8_2) \bigotimes (8_{32}\bigotimes 8_1)
\rightarrow [1_{8_{14}\bigotimes 8_2}\bigotimes 1_{8_{32}\bigotimes 8_1}]
   \bigoplus [O_{8_{14}\bigotimes 8_2}\bigotimes
O'_{8_{32}\bigotimes 8_1}]$, \\
$\bf{(c).}$ $(8_{14}\bigotimes 8_{32}) \bigotimes (8_{1}\bigotimes 8_2)
\rightarrow [1_{8_{14}\bigotimes 8_{32}}\bigotimes 1_{8_{1}\bigotimes 8_2}]
   \bigoplus [O_{8_{14}\bigotimes 8_{32}}\bigotimes
O'_{8_{1}\bigotimes 8_2}]$, \\
$\bf{(d).}$ $8_{14}\bigotimes (8_{32} \bigotimes 8_{1}\bigotimes 8_2)
\rightarrow 1_{8_{14}\bigotimes (8_{32} \bigotimes 8_{1}\bigotimes
8_2)}$, etc., \\
where $O_{8_{14}\bigotimes 8_1}$~($O'_{8_{32}\bigotimes 8_2}$) denotes the 
nonsinglet state formed by $q_1$, $\overline{q}_4$ and 
$g_1$~($q_3$, $\overline{q}_2$ and $g_2$), 
and $O_{8_{14}\bigotimes 8_1} \bigotimes O'_{8_{32}\bigotimes 8_2}$
denotes the color singlet formed by this two nonsinglet states, and so on.
The reduction ways $\bf{(a)}$ and $\bf{(b)}$ lead to the same color 
configurations. The cases $(\bf{c})$ and $(\bf{d})$ give a slightly
different total CR probability from that of the case $\bf{a}$ or $\bf{b}$,
because all of the singlets reduced in $\bf(c)$ and $\bf(d)$
are not color rearranged ones, while the first singlets reduced in
$\bf(a)$ and $\bf(b)$ contribute to the total CR probability. As we can see
in the following, the difference of total CR probability between this
two groups of reduction cases is about 8\%.
Hence the CR probability slightly depends on the reduction way one chooses. 
But our current knowledge of QCD is not enough for us to determine
which the physical reduction way or the physical singlet set is.
As an example, here we discuss the reduction way $\bf{(a)}$. The
corresponding color singlet set $\{|f_2^j>,~j=1,~\cdots,~7\}$ is given by 
\begin{equation}
\label{crea}
\begin{array}{ll}
|f_2^1>=\frac{1}{3\sqrt{8}}|(\Psi_{1i}\Psi_4^i)(\Psi_{3j}\Psi_2^j)Tr(G_1G_2)>,
& |f_2^2>=\frac{1}{\sqrt{80}}|(\Psi_{1i}\Psi_4^i)Tr(G_{32} \{G_1,G_2\})>, \\
|f_2^3>=\frac{1}{12}|(\Psi_{1i}\Psi_4^i)Tr(G_{32} [G_1,G_2])>,
&|f_2^4>=\frac{1}{\sqrt{80}}|(\Psi_{3i}\Psi_2^i)Tr(G_{14} \{G_1,G_2\})>,\\
|f_2^5>=\frac{1}{12}|(\Psi_{3i}\Psi_2^i)Tr(G_{14} [G_1,G_2])>,
&|f_2^6>=\frac{1}{8}|Tr(G_{14}G_{1})Tr(G_{32}G_2)>, \\ 
|f^7>=N|(O_{8_{14}\bigotimes 8_1}\bigotimes O'_{8_{32}\bigotimes 8_2})>, &
\end{array} \end{equation}
where $N$ is the normalization constant for $|f_2^7>$. 
The probabilities of $|f_2^i>$~($i=1,~\cdots,~7$) are defined by
\begin{equation}
P_2^j =\frac{\int\Phi|<f_2^j|H_{c2}|0>|^2 
d\wp_6}{\sigma_2},~~~~~~j=1,~\cdots,~7,
\end{equation}
where $\sigma_2$ is the cross section for the process 
$e^+e^-\rightarrow W^+W^- \rightarrow 
q_1\overline{q}_2q_3\overline{q}_4g_1g_2$, and $H_{c2}$ is given in
eq.~(\ref{c3}). Note that $H_{c2}$ depends on momentum configurations
of partons, so $\sigma_2$ and $P_2^j$ are functions of $\sqrt{S}$ and
the PQCD parameters. In fig.2, we give the probabilities $P_2^j$ at
$\sqrt{S}=170~GeV$. One can notice that they are not sensitive to $Y_{min}$. 

From eq.~(\ref{crea}), we see that
in $|f_2^1>$, $\Psi_{1i}\Psi_4^i$, $\Psi_{3j}\Psi_2^j$ and
$Tr(G_1G_2)$ are three color subsinglets within 
$q_1\overline{q}_4$, $q_3\overline{q}_2$ and $g_1g_2$ respectively. 
Obviously this color separated singlet as we call it is not covered by
the model of Gustafson and H\"{a}kkinen. Its probability is only about
1.3\%. In $|f_2^2>$ and $|f_2^3>$, $\Psi_{1i}\Psi_4^i$ is the color singlet 
within $q_1\overline{q}_4$, while $Tr(G_{32} \{G_1,G_2\})$ 
and $Tr(G_{32} [G_1,G_2])$ are the singlets formed by three octets: one
is symmetric and the other antisymmetric for $g_1$ and $g_2$. The
situations of $|f_2^4>$ and $|f_2^5>$ are similar to those of $|f_2^2>$ and
$|f_2^3>$. In $|f_2^6>$, 
$Tr(G_{14}G_1)$~($Tr(G_{32}G_2)$) is the singlet formed by $q_1$,
$\overline{q}_4$ and $g_1$~($q_3$, $\overline{q}_2$ and $g_2$). 
According to our definition, at $\sqrt{S}=170~GeV$, the total CR
probability  for $e^+e^-\rightarrow W^+W^- \rightarrow 
q_1\overline{q}_2q_3\overline{q}_4g_1g_2$ is about 28\%~(see fig.2)
for the singlet sets ({\bf a}) and ({\bf b}), and about 20\% for ({\bf
c}) and ({\bf d}). Following a similar discussion as we have made for 
$e^+e^-\rightarrow W^+W^- \rightarrow
q_1\overline{q}_2q_3\overline{q}_4g_1$, we see that the meaning of the
probabilities of 
these states are different from what Gustafson and H\"{a}kkinen imply
in their estimation of the CR probabilities. From
their analysis, the total probability of color
rearrangement for this process is $\sim \frac{3}{9}$ at the partonic
level. 

Note that Gustafson and H\"{a}kkinen regard the CR probability at the
partonic level as the upper limit which occurs when the two decay
vertices of $W^+$ and $W^-$ coincide in space-time. The final value
should be decreased by taking into account that two $W$-bosons decay at
different space-time points. But in our approach, the CR probabilities
at the partonic level already include the effects of non-overlapping
decay vertices of two $W$-bosons. This might be the main difference
between these two approaches for the meaning of the CR probability
at the partonic level. 

\section{Summary}
\label{v}
The study of CR in hadronic $W^+W^-$ decays is of significance for both
precisely measuring the mass of $W$ and clarifying the vacuum structure
of QCD. The main goal of this paper is to provide a
strict approach to deriving how many CR singlet sets exist in each final
parton system and calculating the CR probability
at the partonic level. 
To meet this goal, we use the recursive approach to give the invariant
amplitude $M_n$ for the processes $e^+e^-\rightarrow W^+W^-\rightarrow
q_1\overline{q}_2q_3\overline{q}_4+ng$~($n=0,~1,~2$) first; 
Then from $M_n$, we abstract the
corresponding color effective Hamiltonian $H_c$; Finally, $H_c$ is applied to
calculate the CR probability. We find that the CR probabilities 
are $\frac{1}{9}$, $\frac{2}{9}$ and about $20\% \sim 28\%$~(at
$\sqrt{S}=170~GeV$) for $e^+e^-\rightarrow 
W^+W^- \rightarrow q_1\overline{q}_2q_3\overline{q}_4+ng$ 
with $n=0,~1$ and $2$ respectively.

Summarily, the following points should be noted: 
\noindent
\begin{itemize}
\item Our result of the CR probability in PQCD stage is an accurate one 
at the tree level. It already contains the effect caused by different 
space-time intervals between the decay vertex of $W^+$ and that of $W^-$ 
because our approach is a matrix element method. It shows that
the CR probability in PQCD phase is not quite small. 
Our approach to studying CR seems 
different from Sj\"{o}strand and Khoze's. The difference may lie in
the different definition of CR at the partonic level. We defined
CR as the transformation from the original color singlet set to a new one 
where $q_1\overline{q}_4$ and $q_3\overline{q}_2$ belong to different 
color subsinglets, while  
Sj\"{o}strand and Khoze defined that as the "antennae" $\widehat{14}$ and 
$\widehat{32}$ in momentum space. The "antennae"
$\widehat{ij}=\frac{p_i\cdot p_j}{(p_i\cdot k)(p_j \cdot k)}$, where $p_i$,
$p_j$ and $k$ are the momenta of $q_i$, $\overline{q}_j$ and the gluon
radiated from the dipole $q_i\overline{q}_j$. 

~~~In our approach, the color Hamiltonian $H_c$ is abstracted by factorizing 
the invariant amplitude $M$ into the uniform color part multiplying the
momentum one. Then we choose and explore one color singlet set which
contains the CR singlets where $q_1\overline{q}_4$
and $q_3\overline{q}_2$ belong to different color subsinglets. By doing
projection of CR states to $H_c$, i.e., $<f_{CR}|H_c|0>$ where $f_{CR}$
denotes the CR singlet, we derive the CR amplitude, then the cross
section and the probability. In the procedure of calculating the CR
probability, we see that even the color interference terms vanish, there
are also CR contributions which are not of interference origin, which shows
that the color interference contributes to only part of CR. 
In Sj\"{o}strand and Khoze's approach,
all color indices are summed over in $|M|^2$;  
the "antenna" terms of $(\widehat{14})$ and $(\widehat{32})$ imply
CR only arise in the interference sector of
$|M|^2$. The "antennae" can be considered as the sources of further
dipole cascading\cite{dipole}, a parallel description of parton
cascading. As we remember, the parton shower process can be well
described by the GLAP equation where the color 
indices are summed over at each branching point, so GLAP equation is a
probability evolution equation. Another parallel scenario is based on
BFKL evolution equation\cite{bfkl} where the color configuration 
is also smeared. BFKL equation is closely related to the color dipole
model. One great advantage of these approaches
compared with our matrix element method is that they are easy to be
implemented by Monte-Carlo simulation. 
But in these approaches, the other CR sources besides interference 
cannot be fully revealed. 

~~~We are not sure whether the
final event properties predicted from our approach is different from
those from Sj\"{o}strand and Khoze's, because in order to produce final
hadrons from the partonic states one has to apply a phenomenological
model to describe the hadronization process of such color states as
$|f_0^2>$, $|f_1^{3,4}>$ and $|f_2^7>$ etc.. Rigorously say, such
states are beyond the scope of the currently used fragmentation model. 
\item Our results for three lowest order cases correspond to three certain
reduction ways of the color space which lead to one specific color
singlet set for each case that contains the CR states. 
For $q_1\overline{q}_2q_3\overline{q}_4$ or
$q_1\overline{q}_2q_3\overline{q}_4g_1$ case, 
the corresponding total CR probability that
we obtained is the same as what Gustafson and H\"{a}kkinen estimated.
This is a curious coincidence, because for each of these two cases,
only one complete and orthogonal
singlet set leads to CR, i.e., all of the CR states belong to the same
singlet set and orthogonal to each other. 
For $q_1\overline{q}_2q_3\overline{q}_4g_1g_2$ case, there are four
singlet sets that lead to CR. For two of them the total
CR probabilities are the same~($\sim 28\%$ at $\sqrt{S}=170~GeV$),
while for the other two, they are a little smaller~($\sim 20\%$ at
$\sqrt{S}=170~GeV$). The singlet set we choose is one of the former. 
It implies that in general, for a certain final parton system, 
there may exist different CR sets and the corresponding CR probabilities.
We are far from making the physics choice among them. On the other hand, in
this case, some of the CR states are not orthogonal to 
each other since they do not belong to the same singlet set. For
example, one of the rearranged singlets,
$|f'>=\frac{1}{8}|Tr(G_{14}G_2)Tr(G_{32}G_1)>$, belongs to a
singlet set that is different from the set of $|f_2^6>$, so they are not
orthogonal to each other, as we see 
$<f'|f_2^6>=\frac{1}{8}\neq 0$. The probability of a certain CR state is
defined in its own set and is meaningless outside it.
But Gustafson and H\"{a}kkinen do not
discriminate different color singlet sets in their estimation where
the probabilities of non-orthogonal singlets are added up together and 
the total CR probability is more than necessary.
However, our result displays the same tendency that the
gluon emission enlarges the CR probability as predicted by Gustafson
and H\"{a}kkinen.
\item The CR effects observed in the final hadron events still depend
on the hadronization mechanism and the vacuum structure, but they
originate from the color reconnection at 
the partonic level. Only when the CR properties at the partonic level
are made clear is it possible to investigate the nature of
the hadronization mechanism and the vacuum structure.
\end{itemize}

\begin{center}{\bf Figure Captions}\end{center}
\noindent
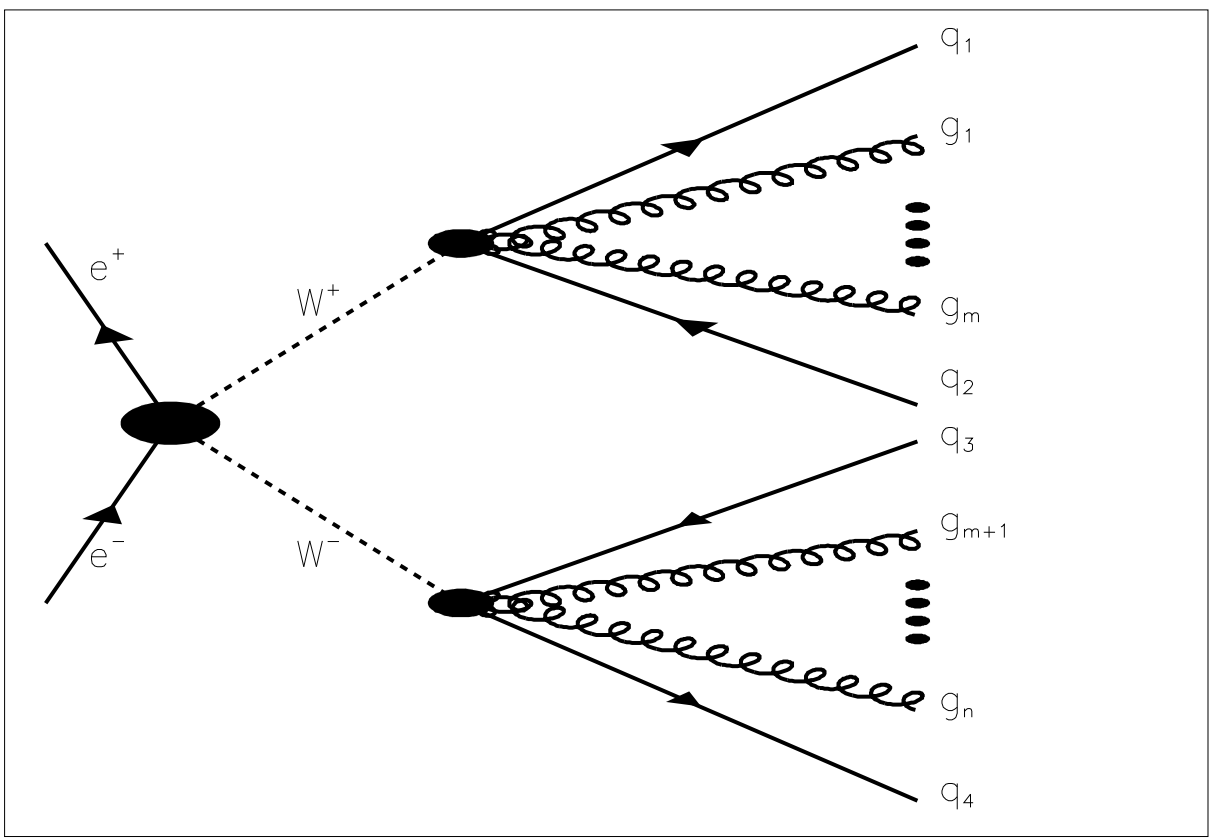
{\bf Fig.1}: $e^+e^-\rightarrow W^+W^-\rightarrow
q_1\overline{q}_2q_3\overline{q}_4+ng$~($n=0,~1,~\cdots$) process. 

\noindent
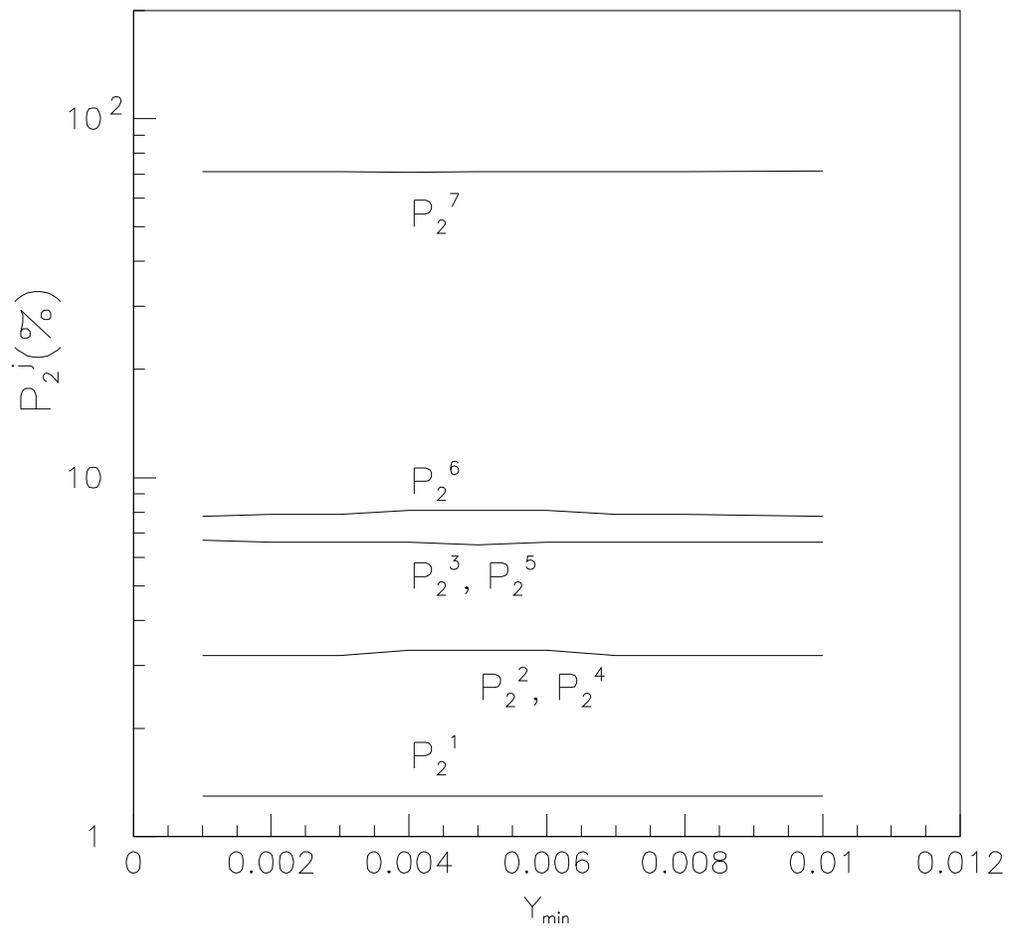
{\bf Fig.2}: The probabilities $P_2^j$ as a function of $Y_{min}$.

\end{document}